\documentclass[journal=achre4,manuscript=review]{achemso}
\usepackage[version=3]{mhchem}

\author{D. Thirumalai}
\email{thirum@umd.edu}
\author{Govardhan Reddy}
\affiliation[University of Maryland]{Biophysics Program, Institute for Physical Science and Technology, Department of Chemistry and Biochemistry, University of Maryland, College Park, MD 20742}
\author{John E. Straub}
\affiliation[Boston University]{Department of Chemistry, Boston University, Boston, MA 02215}

\title{Role of water in Protein Aggregation and Amyloid Polymorphism}
\date{\today}

\begin{document}				

\begin{abstract}

A variety of neurodegenerative diseases are associated with amyloid plaques, which begin as soluble protein oligomers but develop into amyloid fibrils. Our incomplete understanding of this process underscores the need to decipher the principles governing protein aggregation. Mechanisms of \textit{in vivo} amyloid formation involve a number of co-conspirators and complex interactions with membranes. Nevertheless, understanding the biophysical basis of simpler \textit{in vitro} amyloid formation is considered important for discovering ligands that preferentially bind regions harboring amyloidogenic tendencies. The determination of the fibril structure of many peptides has set the stage for probing the dynamics of oligomer formation and amyloid growth through computer simulations. Most experimental and simulation studies, however, have been interpreted largely from the perspective of proteins: the role of solvent has been relatively overlooked in the formation of oligomer assembly to protofilaments and amyloid fibrils. 

In this Account, we provide a perspective on how interactions with water affect folding landscapes of amyloid beta (A$\beta$) monomers, oligomer formation in the A$\beta_{16-22}$  fragment, and protofilament formation in a peptide from yeast prion Sup35. Explicit molecular dynamics simulations of these systems illustrate how water controls the self-assembly of higher order structures, providing a structural basis for understanding the kinetics of oligomer and fibril growth. Simulations show that monomers of A$\beta$ peptides sample a number of compact conformations. The formation of aggregation-prone structures (N$^*$) with a salt bridge, strikingly similar to the structure in the fibril requires overcoming high desolvation barrier. In general, sequences for which N$^*$ structures are not significantly populated are unlikely to aggregate. 

Oligomers and fibrils generally form in two steps. First, water is expelled from the region between peptides rich in hydrophobic residues (for example, A$\beta_{16-22}$), resulting in disordered oligomers. Then the peptides align along a preferred axis to form ordered structures with anti-parallel $\beta$-strand arrangement. The rate-limiting step in the ordered assembly is the rearrangement of the peptides within a confining volume. 

The mechanism of protofilament formation in a polar peptide fragment from the yeast prion, in which the two sheets are packed against each other and create a dry interface, illustrates that water dramatically slows down self-assembly. As the sheets approach each other, two perfectly ordered one-dimensional water wires form. They are stabilized by hydrogen bonds to the amide groups of the polar side chains, resulting in the formation of long-lived metastable structures. Release of trapped water from the pore creates a helically twisted protofilament with a dry interface. Similarly, the driving force for addition of a solvated monomer to a preformed fibril is the release of water; the entropy gain and favorable interpeptide hydrogen bond formation compensate for loss in entropy in the peptides. 

We conclude by offering evidence that a two-step model, similar to that postulated for protein crystallization, must also hold for higher order amyloid structure formation starting from N$^*$. Multiple N$^*$ structures with varying water content results in a number of distinct water-laden polymorphic structures. Water plays multifarious roles in all of these protein aggregations. In predominantly hydrophobic sequences, water accelerates fibril formation. In contrast, water-stabilized metastable intermediates dramatically slow down fibril growth rates in hydrophilic sequences.

\end{abstract}
\newpage

\section{Introduction}

Protein aggregation leading to amyloid fibril formation is linked to a number of neurodegenerative diseases \cite{Shankar08NatMed,Aguzzi09Nature} although in some instances their formation is also beneficial \cite{Maji09Science}. Understanding how misfolded proteins polymerize into ordered fibrils, which universally have a characteristic cross $\beta$-structure \cite{Tycko03Biochemistry}, may be important in our ability to intervene and prevent their formation. The physical basis of protein aggregation involving a cascade of events that drive a monomer to a fibrillar structure  is complicated because of interplay of a number of energy and time scales governing amyloid formation. In addition,  a number of other factors, such as protein concentration, sequence of proteins, and environmental conditions (pH, presence of osmolytes, temperature) affect various kinetic steps in distinct ways, thus making it difficult to describe even \textit{in vitro} protein aggregation in molecular terms. Despite these complexities significant advances have been made,  especially in getting structures of peptide amyloids and models for amyloid fibrils from A$\beta$ and fungal prion proteins. The availability of structures have made it possible to undertake molecular dynamics simulations, which have given insights into the role water plays in oligomer formation as well as assembly and growth of amyloid fibrils.  
 
 It has long been appreciated that water plays a major role in the self-assembly of proteins \cite{Dill90Biochem} in ensuring that hydrophobic residues are (predominantly) sequestered in protein interior. In contrast, the effects of water on protein aggregation is poorly understood. Indeed,  almost all studies (experimental and computer simulations) on amlyoid assembly mechanisms have been largely analyzed using a protein centric perspective.  The situation is further exacerbated by experimental difficulties in directly monitoring water activity during the growth process. Here, we provide a perspective on the role water plays in protein aggregation by synthesizing results from molecular dynamics (MD) simulation studies. Briefly our goals  are: (i) Describe how water-mediated interactions affect the energy landscape of monomers and drive oligomer formation in A$\beta$ peptides.  (ii) The key role water plays in late stages of fibril growth is described by large variations in the sequence dependent mechanism of self-assembly to $\beta$-sheet rich amyloids. (iii) We use results of recent MD simulations and concepts in protein crystallization to provide scenarios for the role water plays in polymorphic amyloid structures.  

\section{Water influences the energy landscape of A$\beta$ monomers}
Although there are several plausible scenarios for the fate of monomer in the conversion to fibrils the process invariably commences by populating misfolded conformations (an ensemble of $N^*$ structures in Fig.~1) by denaturation stress or thermal fluctuations. Thus, the pathways to soluble and mobile oligomer formation and subsequent polymerization depend on the nature of $N^*$, and hence the folding landscape of monomers. Ensemble of $N^*$ (or toplogically related) conformations can collide to populate low order oligomers with differing molecular structures that contain varying  number of water molecules. Once the oligomers exceed a critical size they nucleate and form protofilaments and eventually mature fibrils with differing morphologies (Fig.~1).  Thus, the spectra of the states sampled by the monomers can provide insight into the tendency of specific sequences to form amyloid structures. The relevance of N$^*$ in affecting fibril morphology and growth kinetics suggested in \cite{Thirum03COSB} has been confirmed in a number of studies \cite{Tarus06JACS,Li08JCP,Li10PRL,Shea10BiophysJ}. 

For A$\beta$ peptides and other sequences for which exhaustive MD simulations can be performed it is now established that typically the polypeptide chain samples a large number of conformations belonging to distinct basins, and the aggregation-prone N$^*$structures are separated from the lowest free energy conformations by a free energy barrier. Two extreme scenarios, which follow from the energy landscape perspective of aggregation \cite{Massi01Proteins,Thirum03COSB,Straub10COSB}, can be envisioned.  According to Scenario I, which applies to A$\beta$-peptides and transthyretin, fibril formation requires partial unfolding of
the native state [30] or partial folding of the unfolded state. Both events, which involve crossing free
energy barriers lead to the transient population of an ensemble of assembly-competent structures N$^*$. According to Scenario II, which describes aggregation of mammalian prions \cite{Dima02BiophysJ,Dima04PNAS} the ensemble of N$^*$ structures has a lower free energy than
the structures in the native state ensemble thus making the folded (functional state) state metastable \cite{Baskakov02JBC}. In both scenarios water-mediated interactions are responsible for erecting free energy barriers between the ground state and one of the N$^*$ states. In the case of mammalian prions (PrP) MD simulations \cite{Dima04PNAS} and complementary structural analysis \cite{Dima02BiophysJ} showed that the structured C-terminus must undergo a conformational transition to the more stable N$^*$ structures, which can self-assemble to form self-propagating PrP$^{sc}$ structures.  The need to partially unfold the C-terminal regions results in a substantial barrier between the cellular form of PrP and the aggregation prone N$^*$.

The ensemble of conformations with the lowest free energy in A$\beta_{10-35}$ and the longer A$\beta_{1-40}$ monomer fluctuate\cite{Fawzi07JMB,Sgourakis11JMB,Sgourakis07JMB} among a number of compact structures, whereas in the fibrillar state they adopt a $\beta$-sheet structure. Solid-state NMR-based structural model  of the fibrils of A$\beta_{1-40}$ is characterized by V$^{24}$GSN$^{27}$ turn and intrapeptide salt-bridge between D23 and K28. Such a structural motif, when stacked in parallel,  that satisfies the amyloid self-organization principle \cite{Tarus06JACS,Straub10COSB} according to which fibril stability is enhanced by maximizing the number of hydrophobic and favorable electrostatic interactions (formation of salt-bridges and hydrogen bonds) \cite{Tarus06JACS}. Given that the structural precursors of the fibrils manifest themselves in soluble dynamically fluctuating oligomers, it is natural to expect  that the D23-K28 salt bridge must play an important role in the early events of self-association of  A$\beta$-proteins. Indeed, molecular dynamics simulations of A$\beta_{9-40}$ fibrils suggest that partially solvated D23-K28 salt bridges appear to be arranged as in a one-dimensional ionic crystal \cite{Buchete05JMB}.  However, extensive MD simulations, have shown that the formation of a stable structure with an intact D23-K28 salt bridge and the VGSN turn is highly improbable in the monomer \cite{Tarus06JACS}.  A natural implication is that overcoming the large barrier to desolvation of D23 and K28, which can only occur at finite peptide concentration or by rare flutuations, must be an early event in the formation of higher order structures.

The folding landscape of A$\beta_{10-35}$ can be partitioned into four basins of attraction \cite{Tarus06JACS}. The ensemble of structures with intact salt-bridge, a motif that resembles the one found in the fibril, is rarely populated. There is a broad distribution of compact structures stabilized by a variety of intramolecular interactions. The three most highly populated structures (Fig.~2) are stabilized by solvation of charged residues  and by  hydrophobic interactions in a locally dry environment. Snapshots of the A$\beta_{10-35}$-protein, in which the D23-K28 salt bridge is absent (Fig.~2), show that the two side chains are separated by three and two solvation shells, respectively. Clearly, a stable intramolecular salt bridge can only form
if the interning water molecules can be expelled, which involves overcoming a large desolvation barrier.  Large distance separation between D23 and K28 observed in the first structure in Fig.~2 is due to the interposed side chain of V24 between D23 and K28. This results in a hydrophobic contact between V24 and the aliphatic portion of the K28 side chain. Competition between the electrostatic D23-K28 and the hydrophobic V24-K28 interactions stabilizes the turn in the region V24-N27. The last structure in Fig.~2A shows a D23-K28 water-mediated salt-bridge structure in which one water molecule makes hydrogen bonds with both the D23 and K28 side chains. The lifetime of hydrogen bonds of the solvated D23 and K28 is nearly three times ($\sim$ 2.4 ps) as long as bulk water ($\sim$ 0.8 ps). Thus,  the side chains of D23 and K28 make stronger contacts with water than water with itself, indicating that the desolvation of D23 and K28 is an activated process. The barrier can be reduced by creating monomers containing preformed D23-K28 salt bridge. Indeed, experiments show that aggregation of monomers containing a lactam-bridge between D23-K28  aggregate $\approx1000$ times faster than the wild type\cite{Meredith05Biochem}. This finding has been rationalized in terms of a reduction in the free energy barrier between low free energy structures without D23-K28 salt bridge and N$^*$ structures (ones in which these residues are in proximity) in chemically linked monomers\cite{Reddy09JPC}. 

\section{Dynamics of Oligomer Formation}
The first MD study on interacting peptides \cite{Klimov03Structure} focused on the mechanism of assembly of  peptide fragment KLVFFAE [A$\beta_{16-22}$]$_n$ ($n$ = 2 and 3), which contains the central hydrophobic cluster LVFFA (CHC)  flanked by the N-terminal positively charged  residue (Lys)  and the C-terminal negatively charged residue (Glu).   The peptides form antiparallel $\beta$-sheet structure in the fibril as assessed by solid state NMR and molecular dynamics simulations. Somewhat surprisingly,  MD simulations showed that even in a trimer the peptides, which are unstructured as  monomers \cite{Klimov03Structure}, are extended and arranged in antiparallel fashion. Such an arrangement ensures formation of the largest number of inter peptide salt-bridges in addition to maximizing the number of hydrophobic contacts between the peptides, which accords well with the amyloid-organization principle. Thus, the ordered structure, which undergoes substantial conformational fluctuations because of finite size, should be viewed as a "nematic" droplet in which the strands are aligned along a common director. Explicit mapping using  analysis of MD trajectories showed that the energy landscape (A$\beta_{16-22}$)$_2$) has nearly six minima \cite{Garcia06JACS} including one in which the peptides are antiparallel to each other.  However, the number of minima decrease as $n$ increases and approached a critical size \cite{Li07PNAS}.

The mechanism of oligomer formation revealed that the salt-bridges gives rise to orientational specificity, which renders the antiparallel arrangement stable. However, the driving force for oligomerization, which initially produces an ensemble of disordered aggregates, is the hydrophobic interaction between various residues in the CHC.  The early formation of disordered structures in LVFFAE implies that water must be expelled relatively quickly upon interaction between the peptides. It was found that at very early stages the number of water molecules is substantially reduced from the crevices between the peptides which implies that  the ordered  nematic droplet, which is coated on the outside with water, is essentially dry.  The first MD studies\cite{Klimov03Structure} showed that expulsion of water in sequences with a large number of bulky hydrophobic residues must be an early event, and hence cannot be the rate limiting step in the ordered assembly of such peptides.

The growth dynamics of oligomers of A$\beta_{16-22}$ peptides further showed that water is not present in the interior. Simulations of the reaction (A$\beta_{16-22}$)$_{n-1}$ + A$\beta_{16-22} \leftrightarrow$ (A$\beta_{16-22}$)$_n$ done by adding an unstructured solvated monomer to a preformed oligomer it was shown that the monomer adds onto the larger particle by a dock-lock mechanism \cite{Li07PNAS}. In the first docking step the solvated monomer attaches to the oligomer rapidly by essentially a diffusive process. In the much slower lock step the peptide undergoes conformational transitions from a random coil  to a $\beta$-strand conformation, and adopts a conformation that is commensurate with the structures in the nematic droplet.  Interestingly, the interactions that stabilize the larger oligomers also do not involve water molecules. In addition, there are very few stable hydrogen bonds that persist between the peptides, which implies that the higher order oligomer structures are stabilized largely by interpeptide side-chain contacts. As is the case  for trimers, the antiparallel orientation is guaranteed by the formation of the salt bridge between K16 from one peptide and E22 from another.  Taken together, these results show that the driving force for oligomerization is the favorable interpeptide association between residues belonging to the CHC.   Although the role of side chains is a major determining factor in oligomer formation in all peptides it should be stressed that  expulsion of water from the interior of oligomers in the early stages is highly sequence dependent.   

\section{Water release promotes protofilament formation and amyloid fibril growth}

Studies of protein crystallization\cite{Vekilov_soft_2010} remind us that a major driving force for crystal formation is the release of water molecules from the hydration layer upon formation of contacts between protein molecules. A number of experimental, simulation, and theoretical studies of proteins, which crystallize with intact folded structures have shown that even in cases when enthalpy gain upon crystallization is small, it is more than compensated by depletion of water molecules around the proteins. Crystallization results in a loss in translational and rotational entropy and the vibrational degrees of freedom associated with the ordered structure only partly compensate for the loss. However, the total entropy change, $S_T$, ( = $\Delta S_{protein} + \Delta S_{water}$ where $\Delta S_{protein}$ and $\Delta S_{water}$ are the changes in protein and water entropy, respectively) is positive, and is explained by water release mechanism.    Water is structured around the surface of folded proteins with the  thickness of  hydration layer being $\sim$ 7\AA \ (Fig.~1). Water molecules in the structured layer are  in dynamic equilibrium with the bulk water.  Upon crystallization, the structured water (typically $\sim$ (5-30)) around the protein is released into the bulk, which contributes to an increase in $S_T$, and has been suggested as a major thermodynamic driving force for protein crystallization \cite{Vekilov_soft_2010}.  Thus, even if the polymerization process is endothermic, increase in $S_T$ can drive gelation and crystallization. Self-assembly of  Tobacco Mosaic Virus (TMV) \cite{Lauffer_Biochem_1965,Lauffer_Biochem_1969} provided an early example of water-release mechanism.  The endothermic polymerization reaction\cite{Lauffer_Nature_1958} involving TMV goes to completion by the water release mechanism leading to an increase in $S_T$ of the system. Using the quartz spring balance experiments\cite{Lauffer_Biochem_1969}, it is demonstrated that 96 moles of hydrated water is released per mole of the TMV protein trimer which is shown to be sufficient to increase the overall entropy of the system and drive the polymerization reaction. There are a number of experimental studies involving protein crystallization, which are nearly quantitatively explained by increase in $\Delta S_{water}$. 

Although not discussed explicity, expulsion of water leading to increase in solvent entropy, which has been observed in oligomer formation of A$\beta$-peptides \cite{Klimov03Structure, Li07PNAS,Krone09JACS,Reddy_PNAS_2009}, is also a major driving force for fibril formation. Here, it is important to consider both sequence effects and account for conformational changes that occur. For example, in the case of A$\beta_{16-22}$ the random coil structure (small size) expands to form $\beta$-strand (larger size), which is unfavorable not only because it involves solvent exposure of hydrophobic residues but also results in reduction in conformational entropy.  In this case both release of water and favorable side chain contacts stabilize the oligomers.  Aggregation  of Sup35 yeast prion protein \cite{Lindquist_nature_05}, the N terminal region of the protein rich in polar side groups (glutamines and asparagines) participate in the formation of collapsed disordered aggregates. Simulations have shown, somewhat surprisingly, that water is a poor solvent (we adopt the terminology used in polymer physics) for the polypeptide backbone and collapsed disordered polyglutamine chains are thermodynamically favored \cite{Pappu_jacs_2005,Pappu_JMB_2008,Pappu_PNAS_2009}. To decrease the interfacial tension between the water and the backbone secondary amides, polyglutamine becomes compact with the side chain amides solvating the backbone amides.  Thus, different driving forces are involved in the initial collapse of protein molecules leading to disordered oligomers and the mechanism depends on the protein sequence. The common  universal driving force is predominantly the release of the structured water \cite{Perutz_PNAS_2002} around the protein into the bulk as oligomers, protofilament, and fibrils form. Because the strength of interpeptide interactions and solvent mediated forces are sequence-dependent, time scales for fibril formation also can vary greatly depending on the sequence even under identical external conditions \cite{Reddy_PNAS_2010}.


Two recent simulations on the growth of fibrils (assumed to occur by incorporating one monomer at a time \cite{Reddy_PNAS_2009}) and self assembly of protofilament  \cite{Reddy_PNAS_2010} vividly illustrate water release as a key factor.  Addition of a Sup35 peptide (GNNQQNY)  to an amyloid fibril reveal \cite{Reddy_PNAS_2009} that the release of the hydrating water molecules into the bulk and peptide addition to the fibril occur simultaneously (Fig.~3). The number of water molecules, $N_W(t)$  decreases as the solvated monomer interacts with the underlying fibril lattice (Fig.~3A).  As the locking reaction progresses, water molecules in the vicinity of the monomer in the fibril that are closest to the solvated monomer are released (Fig.~3B). Comparison of the growth dynamics associated with the A$\beta$ peptides and Sup35 shows that the dehydration process is dynamically more cooperative for the polar sequence \cite{Reddy_PNAS_2009}, which emphasizes the role of sequence discussed above.  Fluctuations in the number of water molecules coincide with the locking events (Fig.~3). The largest fluctuations in the number of water molecules near the locking monomer, $N_W^L(t)$ and the solvent-exposed monomer in the fibril, $N_W^F(t)$ occur precisely when the monomer completely locks onto the crystal cooperatively (Fig.~3 A). The coincidence of the locking step and dehydration is also reflected in the sharp, decrease in the water content in the zipper region of the Sup35 crystal (Fig.~3B), which occurs in two well-separated stages. The number of water molecules, $N_W^Z(t)$ decreases abruptly from 8 to 2 as the docking is initiated, and finally goes to zero as the locking process is complete (Fig.~3B). These observations show that dehydration leading to release of "bound" water, resulting in the formation of the dry zipper region must be taken into account in estimating free energy changes that occur upon amyloid fibril growth. 

In a recent study we predicted that there must be large variations (exceeding a factor of over one thousand) in the time needed for self-assembly of protofilaments between hydrophobic and polar sequences because of the entirely different roles water plays in their formation \cite{Reddy_PNAS_2010}. The barrier to the release of bound water around the polar residues  should be high due to the favored interactions between the polar side chains and water compared to  hydrophobic side chains.  As a consequence, protofilaments comprised of polar sequences must take much longer to form than ones made of hydrophopbic residues. These expectations were borne out in MD simulations \cite{Reddy_PNAS_2010} contrasting the role of water in the protofilament formation from peptides with polar and non-polar residues.  Water forms spontaneously meta-stable ordered one-dimensional wires in the pores of the protofilaments  during the assembly of the $\beta$-sheets of GNNQQNY preventing the sheets to associate completely. The water wires are stabilized by the hydrogen bonds with the amide groups in the side chains of asparagines and glutamines and delay the protofilament formation. The gain in entropy due to water release can be obtained by comparing the difference in the energies (obtained from MD simulations\cite{Reddy_PNAS_2010}) between the metastable and stable structures. We find that entropy gains per water molecule released is $\approx$ 6 cal/mole.K, which is similar to the value estimated from protein crystallization experiments\cite{Vekilov_soft_2010}. 

There are predominantly two major routes to assembly of $\beta$-sheets (Fig.~4). In one of them spontaneously formed nearly perfectly ordered one-dimensional water from the pore is released into the bulk resulting in  $\beta$-sheet association and protofilament formation. Alternatively,  when fluctuations lead to  misalignment in the orientation of the $\beta$-sheets, water release occurs by leakage through the sides.  In such a pathway the sheets are packed against each other with orientational defect, and could represent one of the polymorphic structures.  In contrast, the assembly of the $\beta$-sheets of GGVVIA, rich in hydrophobic groups occurs rapidly, and  the water in between the sheets is eliminated concurrently as the $\beta$-sheets associate with one another \cite{Reddy_PNAS_2010}.  The contrasting behavior observed in the protofilament assembly observed in hydrophobic and polar sequences illustrates the distinct role of water. In the former case the driving force forming a protofilament with a dry interior is the hydrophobic interactions. However, if  the amyloid forming sequence is hydrophilic then water release serves as a substantial driving force. In this case water is a surrogate hydrogen bond former, upon release of the trapped water, protofilament assembly is completed.

Similarly, simulations \cite{Krone09JACS} of association between preformed $\beta$-sheets in A$\beta_{16-22}$ showed that in some of the trajectories water is expelled early before assembly. In other trajectories, the two processes are observed to be coincident. The predominant interactions that mediate protofilament formation are hydrophobic with interactions involving Phe playing a major role, as was previously shown in the context of oligomer formation. In both cases water release provides the needed impetus for self-assembly. The simulations also rationalize experiments\cite{Gai_JPC_2009}, which showed that the rate of fibril formation increases significantly on reducing the hydration of aggregating peptide molecules. It was found that aggregation rate of A$\beta_{16-22}$ is largest when stabilized in reverse AOT micells containing the least amount of free water molecules.

\section{Is Water part of  polymorphic structures?}
The consequences of misfolding to multiple conformations  with subsequent aggregation into distinct infectious states with differing phenotypes (the so called strain phenomenon) has been established in prion disorders and A$\beta$ peptides. Originally found in the context of wasting diseases  and mammalian prions, strain phenotypes, which grow from the same protein but lead to different heritable states, are found even in peptide fibrils and amyloids grown from A$\beta$-peptides. In general, amyloid fibrils show polymorphism both  in the  mature structure\cite{Petkova_Science_2005,Petkova_Biochem_2006,Sawaya_Nature_2007,Paravastu_PNAS_2008}, and is also manifested in protofilaments \cite{Sawaya_Nature_2007}. Various structures differ in side chain packing, water content, hydrogen bond networks (Fig.~1) or in the quaternary structure \cite{Petkova_Biochem_2006,Paravastu_PNAS_2008,Petkova_Science_2005}. Polymorphism in amyloid fibrils also forms the basis of strain phenomena in prion protein\cite{Toyama_Nature_2007}. A single prion protein with multiple infectious conformations, one for each strain, gives rise to distinct phenotypes and is also inheritable \cite{Tessier_NSB_09}. Although polymorphism is widely observed in amyloid fibrils, the biophysical basis for their formation is lacking. 

It is likely that trapped water molecules is part of the observed polymorphic structures. The rationale for such a suggestion is based on the energy landscape perspective of protein aggregation (Fig.~1), which provides a plausible connection to the strain phenotypes that have been extensively studied especially in yeast prion biology.  At what stage of the growth of fibrils is a particular strain `encoded' in the structure? The suggestion that the N$^*$ structures are aggregation prone implies that the strain phenotypes may be encoded in the monomer structures or low order oligomers. We speculate that the various N$^*$ structures can form oligomers with different structures, which can subsequently lead to  structurally distinct fibrils.  

It is also clear from MD simulations \cite{Klimov03Structure,Li07PNAS,Tarus06JACS} that the pre-nucleus structures, which we propose are candidates for encoding polymorphism, are water-laden. Hence, it follows that the distinct mature fibrils must contain discrete number of water molecules. A number of studies provide evidence for our proposal. Formation of water channels near the salt-bridge (D23-K28) has been observed in simulations \cite{Buchete05JMB} of a solid-state NMR-derived structural model. In the resulting structure, which is a variant of the one proposed using experimental constraints, the buried salt bridge between D23 and K28 are arranged in a periodic manner along the fibril axis. Confined water molecules solvate the salt bridge, which is interestingly reminiscent of high free energy conformations sampled by A$\beta_{10-35}$ monomer (see the last structure in Fig.~2A). More recently, a different morphology for A$\beta_{1-40}$ has been  proposed using 2D IR spectroscopy \cite{Hochstrasser_PNAS_2009}.  It was found that water molecules (roughly 1.2 per monomer) are trapped in A$\beta_{1-40}$ fibrils.  However, the finding that water molecules are trapped in the hydrophobic pocket (L17, V18, L34, and V36) that interact with the amide backbone of L17 and L34 is a surprise. There are two possible explanations for these findings. If mobile water molecules are not part of the fully mature fibrils, it is likely that the fibril structures with trapped waters are metastable. If this were the case then we could argue that on much longer time scales the trapped water molecules would migrate closer to the charged residues and populate a structure similar to that found in MD simulations \cite{Buchete05JMB}.  Alternatively, it is possible that these structures represent a distinct polymorphic fibril structure. We surmise that other proposed structures for A$\beta$-peptides must contain  discrete number of water molecules trapped in the fibril interior, and hence must be part of amyloid polymorphism.   

The scenarios of fibril formation (Fig.~1) and lessons from protein crystallization provide a physical picture for polymorphism. It is firmly believed that crystallization generally and protein crystallization  in particular occurs in two steps \cite{Vekilov_soft_2010,tenWolde97Science}. In the first step, fluctuations produce droplets that are  rich in proteins leading to structures that are disordered.  In the second stage rearrangement of the structures within the droplet produces ordered structures which grow by incorporating one monomer at a time.  Globally a similar mechanism qualitatively explains amyloid formation (Fig.~1) \cite{Dima02Prot,Nguyen04PNAS,Pellarin06JMB,Auer07HFSP,Bellesia09JCP,Liang_JACS_2010}. The first step involves formation of disordered oligomers which produces regions that are protein rich droplets.   In contrast to protein crystallization in which proteins are folded, the N$^*$ structures in the droplet could contain varying number of water molecules that may be embedded in the mature fibrils. Once the droplet size becomes large enough (by collisions with smaller droplets or by monomer addition) they produce distinct fibrillar structures, which differ not only in inter protein interactions but also in the content of water. The two-step growth mechanism, which is reminiscent of nucleated conformational conversion picture \cite{Serio00Science}, differs from the traditional nucleation mechanism because growth occurs within the liquid-like disordered droplets that are protein rich. As a result, morphologies which nucleate more frequently dominate  fibril formation rather than ones which are thermodynamically more stable. Thus, the dominant fibril morphology emerges from those N$^*$ structures that minimize surface energies in the protein rich droplets. It also follows that distinct strain formation might be under kinetic control \cite{Pellarin_JACS_2010} .   

\section{Conclusions}
Naturally occurring peptides and proteins that form $\beta$-amyloids are wonderful systems that can be used to study self-assembly of higher order structures and hydration dynamics. Extrapolation of such biophysical studies to what transpires physiologically is often fraught with difficulties. For example, damage to synapses in Alzheimer's disease is not caused solely by oligomers of A$\beta$-peptides whose production  is a complicated process involving other enzymes. There are other culprits  (one or more kinases)  whose interaction with A$\beta$-oligomers apparently play a significant role in synapse impairment. Thus, bridging the gap between \textit{in vitro} and \textit{in vivo} studies involve protein-protein recognition, which could also be mediated by water in different ways.  

From a biophysical perspective characterizing the nature of fluctuations that promote regions that are rich in N$^*$ peptides (Fig.~1), which is a precursor to growth of ordered $\beta$-amyloids, as the protein concentration is lowered from $\sim$ mM (used in computer simulations) to $\sim \mu$M (needed to grow fibrils in the laboratory) to $\sim$ nM (found in \textit{in vivo}) conditions is a challenging problem. Molecular dynamics simulations that probe water-mediated interactions and the associated dynamics  in mesoscale droplets containing a number of amyloidogenic species will go a long way in our ability to describe self-assembly of $\beta$-amyloids. In such confined spaces water alters both hydrophobic and electrostatic interactions substantially \cite{Vaithee05JACS,Vaithee08PNAS}. Indeed, amyloid fibrils, which can be  can be pictured as water filled nanotubes \cite{Perutz_PNAS_2002}, are great systems to probe  the properties of confined water. In these systems hydrophobic or hydrophilic environment for the confined water can be controlled using mutations, which naturally changes water density in the pores. It is clear from this brief perspective that many facets of amyloid growth, driven by context dependent interactions involving water and by the peculiarities of confined water, remain to be explored using carefully planned simulations and experiments. Finally, it should be noted that lessons from such simulations can and should be incorporated into simpler models and theories as summarized in recent reviews \cite{Straub10COSB,Shea11COSB,Straub11ARPC}.   

\textbf{Acknowledgements} This work was supported by a generous grant
from the National Institutes of Health (GM076688-08).


\providecommand*\mcitethebibliography{\thebibliography}
\csname @ifundefined\endcsname{endmcitethebibliography}
  {\let\endmcitethebibliography\endthebibliography}{}

\newpage

{\bf Figure Captions:}
\bigskip

Fig.~1: Schematic of protein aggregation mechanisms leading to polymorphic fibrils. On the left are solvated peptides. Water in the hydration layer is in red and the bulk water in blue. Even isolated monomers sample aggregation prone conformations, N$^*$, which are coated with varying number of water molecules. The peptides with N$^*$ conformations aggregate to form disordered protein rich droplets. A major  driving force for  aggregation is the release of  water molecules in the hydration layer into the bulk, which facilitates fibril formation entropically favorable. The structured protein aggregates nucleate from the protein rich droplet to form protofilaments, which further self-assembles  to form a variety of mature amyloid fibrils. In some of the polymorphic structures discrete number of water molecules are confined in the fibril. 

\bigskip

Fig.~2: Folding landscape of A$\beta_{10-35}$ monomers.  (A) Low free energy conformations  in the which  D23 and K28 amino acids, which forms a salt bridge in the fibril, are separated by three, two and one water solvation shells respectively (from top to bottom).  The backbone oxygen and nitrogen atoms are in red and blue, respectively. The positively and negatively charged, polar, and hydrophobic residues are colored  blue, red, purple, and green, respectively. Water molecules around D23 and K28 are in cyan, while water molecules which separate the two residues are shown in yellow. Hydrogen bonds are shown as black dashed lines. (B) Hairpin-like conformation of the A$\beta_{10-35}$ monomer which has a topologically similar structure as the peptide structure in the A$\beta$ fibrils. The D23-K28 salt bridge is solvated by the water molecules. The driving force for the formation of hairpin-like conformation is the interaction between the hydrophobic residues in the N and C termini shown in red and green  respectively. 

 \bigskip

Fig.~3: Water release in fibril growth. (A) Variation in the number of water molecules (red) within 3.5\AA \ of the peptide from Sup35, which docks and locks onto the fibril as a function of time. Time-dependent changes (green)  in the number of water molecules  in the neighborhood of the fibril monomer onto which the solvated peptide docks.  (B) Release of water molecules in the zipper region of the fibril occurs in  two stages. In the first stage, water is eliminated rapidly as the peptide docks onto the fibril, while in the second stage, the last two water molecules are squeezed out with the concurrent formation of the protofilament with a dry interior (structure on the right).

 \bigskip

Fig.~4: Water molecules play a central role in the association kinetics of two sheets formed from peptides rich in amino acids with polar side chains. In the association process starting from a fully solvated pore (structure on the left) trapped water molecules between the protofilaments form ordered water wires (top middle structure). If the sheets misalign confined water molecules are  disordered.  Release  of  trapped water molecules  results in  protofilament formation (structure on the right). In the upper pathway the water molecules in the wire file out in orderly fashion whereas in the bottom pathway water escapes from the crevice on the sides of the protofilament.

\newpage

\newpage
\begin{figure}[ht]
\includegraphics[width=7in]{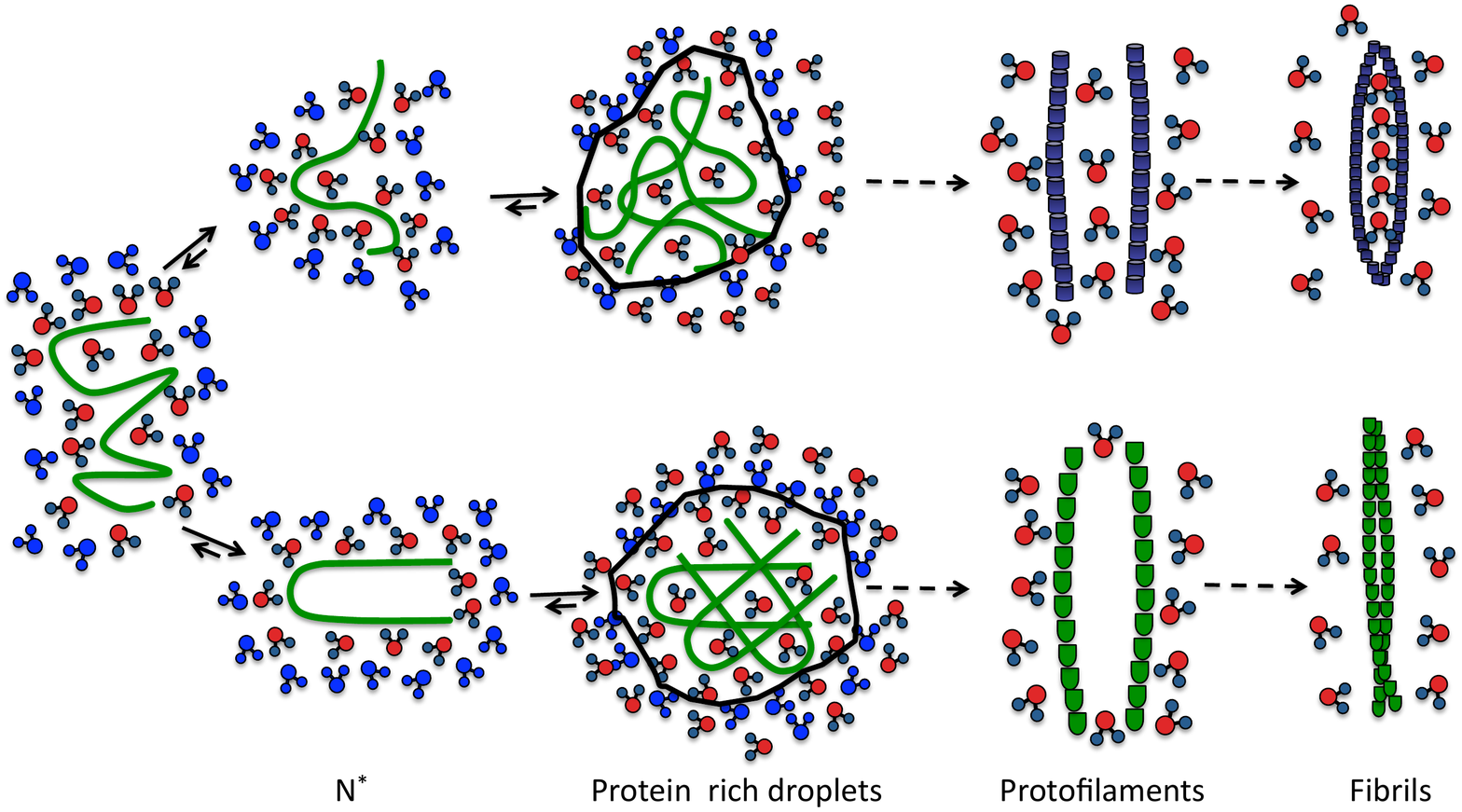}
\label{fig1}
\end{figure}

\begin{figure}[ht]
\includegraphics[width=3in]{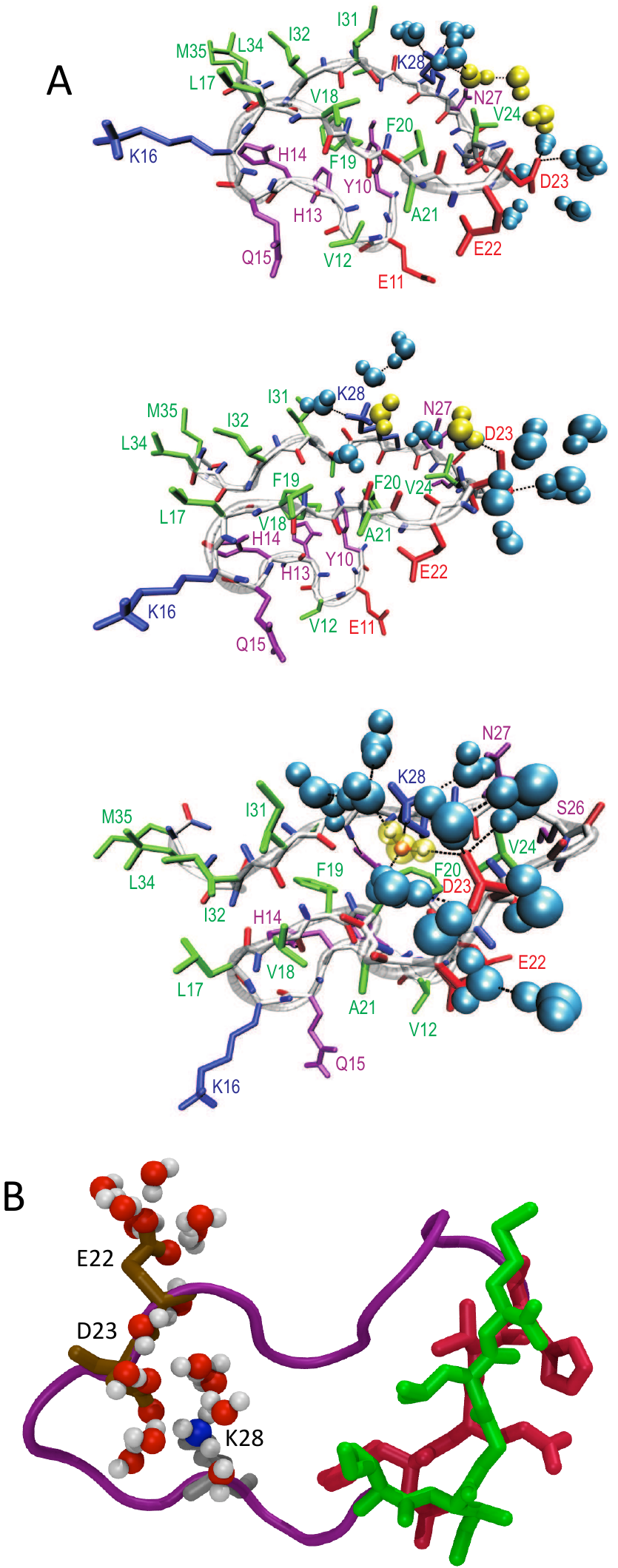}
\label{fig2}
\end{figure}

\begin{figure}[ht]
\vspace{0.6 in}
\includegraphics[width=5in]{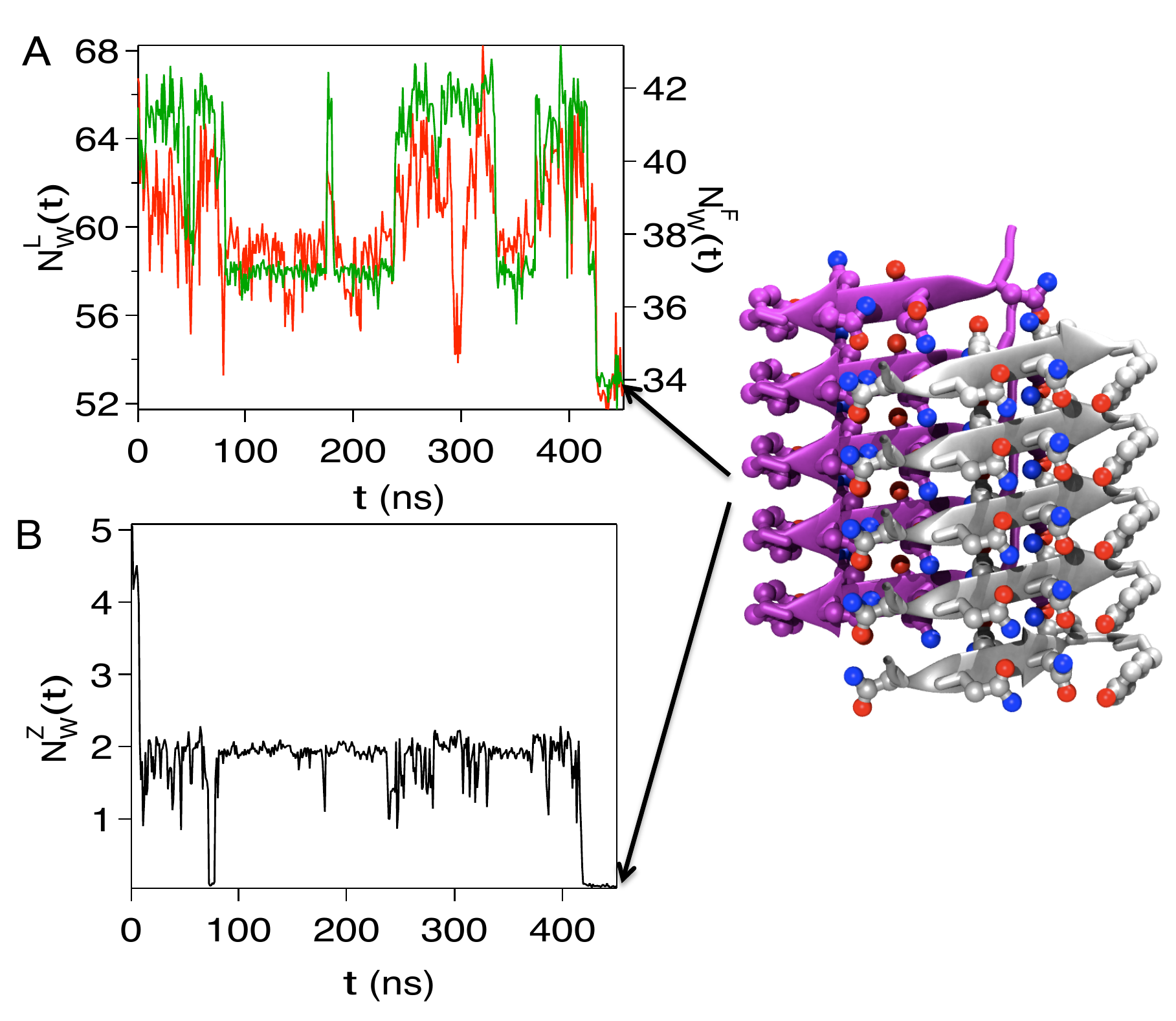}
\label{fig3}
\end{figure}

\begin{figure}[ht]
\includegraphics[width=5in]{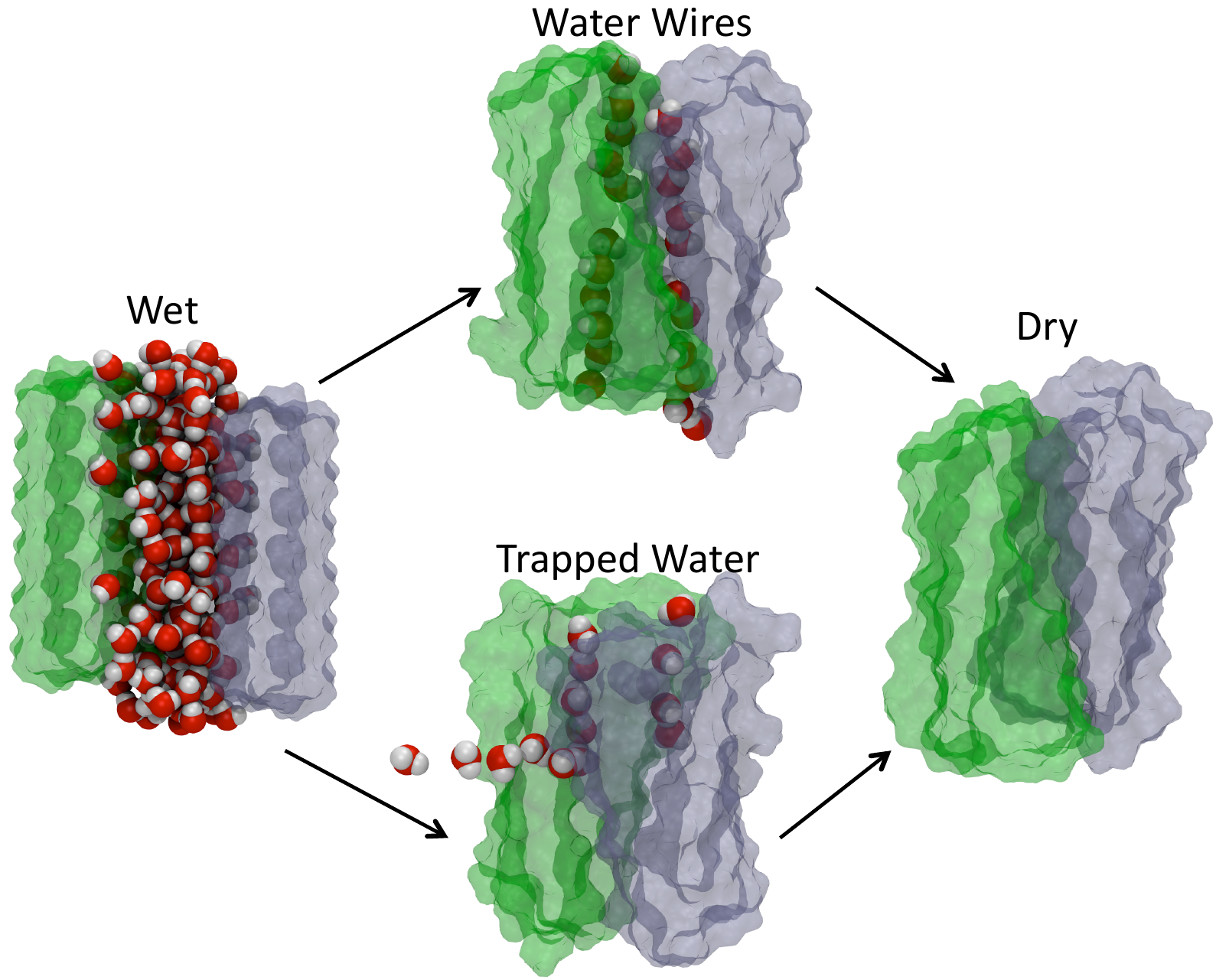}
\label{fig4}
\end{figure}

\end{document}